# Estimation of the integral role of Intershell correlations in heavy atoms

M. Ya. Amusia[1, 2][1] and L. V. Chernysheva[2]

[1]*The Racah Institute of Physics, the Hebrew University, Jerusalem 91904, Israel*
[2]*A. F. Ioffe Physical-Technical Institute, St. Petersburg 194021, Russian Federation*

**Abstract:**
We have calculated partial contributions of different atomic subshells to the total dipole sum rule in the frame of the random phase approximation with exchange (RPAE) and found that they are essentially different from the numbers of electrons in respective subshells. This difference manifests the strength of the intershell interaction, due to which some partial contributions are much bigger while the other are considerably smaller than the numbers of electrons in the respective shells. Particularly impressive is the growth of contribution of the outer among f and d subshells while all other are usually losers, the biggest of which are the s-subshells. Concrete calculations done for Ar, Pd, Xe, and Ra atoms. Comparison with existing experimental data is uninformative since to obtain the absolute values of the cross-sections one usually normalizes them, assuming that the dipole partial sum rule is valid with reasonable accuracy.

PACS numbers: 32.70.Cs, 32.80.-t, 32.80.Fb

**1**. In this Letter, we demonstrate that by studying the dipole sum rule that correspond to a given subshell one obtains important information on the interaction between electrons of this subshell and the other atomic electrons.

W. Thomas, F. Reiche, and W. Kuhn have discovered the Dipole sum rule almost a century ago [1, 2]. It connects oscillator strengths of discrete $f_k$ transitions, dipole non-relativistic photoabsorption cross-section $\sigma(\omega)$ as a function of incoming photon frequency $\omega$ and the number of electrons $N$ in the system, e.g. atom that absorb photons, by the following relation [3][2]:

$$\sum_{\text{All } k} f_k + \frac{c}{2\pi^2}\int_I^\infty \sigma(\omega)d\omega \equiv S = N. \qquad (1)$$

Here $I$ is the ionization potential and $c$ is the speed of light. The relation (1) is a strict theoretical statement in non-relativistic approximation, but experimentally observed cross-sections and oscillator strengths has admixtures of non-dipole contributions and of a whole variety of relativistic corrections. These corrections grow rapidly with increase of $\omega$ and become dominative in $\sigma(\omega)$ at relativistic energy $\omega \geq c^2$. Fortunately, for atoms and solid bodies photon energies $I \leq \omega \ll c^2$ saturate the integral in (1), so the role of dipole contributions is most important. To determine experimentally the absolute values of the photoionization cross-section is a hard task, so instead one normalizes the accurately measured relative values of $\sigma(\omega)$ using (1). This is a regular procedure not only for atoms, but also for more complex objects, e.g. such, as endohedrals and fullerenes [4].

---

[1] amusia@vms.huji.ac.il
[2] We employ the atomic system of units $m = e = \hbar = 1$. Here $m$ is the electron mass; $e$ is its charge and $\hbar$ is the Planck constant.

It exist a widely distributed believe that an equation, similar to (1), although approximate, is valid for partial subshell contributions [5]:

$$\sum_{\text{All } k_i} f_{k_i} + \frac{c}{2\pi^2} \int_{I_i}^{\infty} \sigma_i(\omega) d\omega \equiv S_i \approx N_i. \qquad (2)$$

Here the oscillator strength $f_{k_i}, \sigma_i(\omega), I_i$ and $N_i$ are, respectively, the discrete excitations oscillator strengths, photoionization cross-section, ionization potential and total number of electrons in the $i^{th}$ subshell. A usual assumption is that relation (2) is accurate enough to attribute absolute values to the measured partial cross-sections.

Known almost half a century, the random phase approximation with exchange (RPAE) gives very good results in description of partial photoionization cross-sections, angular distributions of photoelectrons and spin polarization parameters [5]. An interesting feature of this approximation is the fact that the dipole sum rule (1) is precisely valid in its frame. In this Letter we investigate the partial sum rules in RPAE frame and demonstrate that, unexpectedly, for multi-electron $f$ and $d$ subshells the values $S_f$ and $S_d$ are considerably bigger than the respective $N_f = 14$ and $N_d = 10$. It means that for other subshells the inequalities hold $S_{s,p} < N_{s,p}$ where $N_s = 2$ and $N_p = 6$ [3], in generally holds. Objects of concrete calculations are Ar, Pd, Xe, and Ra atoms. We employ the one-electron Hartree-Fock (HF) approximation, in both length and velocity forms of the operator that describes the photon-electron interaction, denoted by lower indexes $L$ and $\nabla$. In HF the relation (1) is essentially violated. We take into account the multi-electron correlations in the RPAE frame, where Eq. (1) is valid.

**2.** The necessary details about HF and RPAE equation and their solutions one can find in [6, 7]. Here we present only important definitions and the main points of calculation procedures. The HF equation for multi-electron atoms looks like

$$-\frac{\Delta}{2}\phi_j(x) - \frac{Z}{r}\phi_j(x) + \sum_{k=1}^{N} \int \phi_k^*(x') \frac{dx'}{|\mathbf{r}-\mathbf{r}'|} \left[\phi_k(x')\phi_j(x) - \phi_j(x')\phi_k(x)\right] = E_j \phi_j(x). \qquad (3)$$

Here Z is the nuclear charge, $\phi_j(x)$ is the one-electron wave function, $x \equiv \vec{r}, \vec{\sigma}$ are the electron coordinate and spin variables, respectively $E_j$ is the one-electron HF energy; the summation is performed over all occupied electron states $N$.

The oscillator strength is determined by the square module of dipole matrix elements in the length $\vec{\varepsilon}\vec{r}$ or velocity $\vec{\varepsilon}\vec{\nabla}$ forms, calculated between HF wave functions (3) of the electron that undergoes transition from the initial state $i$ to the final $f$ due to photon absorption:

---

[3] All considered here atoms have only closed subshells.

$$d_{if}^L = \omega_{if} \int \phi_i^*(x)(\vec{\varepsilon}\vec{r})\phi_f(x)dx, \quad d_{if}^\nabla = \int \phi_i^*(x)(\vec{\varepsilon}\vec{\nabla})\phi_f(x)dx, \quad \omega_{if} \equiv E_f - E_i \qquad (4)$$

The following expression determines the oscillator strength of a one-electron transition $i \to f$:

$$f_{if}^{L,\nabla} = \frac{2}{\omega_{if}} \left| d_{if}^{L,\nabla} \right|^2, \qquad (5)$$

Similar to (5) expression is valid for continuous spectrum excitations that is connected to the photoionization cross-section of the $i$ subshell by the following relation

$$\sigma_i^{L,\nabla}(\omega) = \frac{2\pi^2}{c} f_{iE_{(i)}}^{L,\nabla}, \quad E_{(i)} = \omega - I_i. \qquad (6)$$

Relations similar to (5) and (6) give the oscillator strengths and photoionization cross-section in RPAE, if one substitutes the HF matrix elements $d_{if}^{L,\nabla}$ by solutions of RPAE equations

$$\langle i|D(\omega)|f\rangle = \langle i|d^{r,\nabla}|f\rangle + \left( \sum_{v \leq F, v' > F} - \sum_{v > F, v' \leq F} \right) \frac{\langle v'|D(\omega)|v\rangle \langle vi|V|v'f - fv'\rangle}{[\omega - E_{v'} + E_v \pm i\delta)]}. \qquad (7)$$

Here $V$ denotes the Coulomb inter-electron interaction, sums over $v \leq F$ include occupied one-electron states, while sums over $v > F$ include excited discrete levels and integration over continuous excitation energies. In the denominator the sign $\pm$ means + for $v'$ vacant and – for $v'$ occupied one-electron states, respectively. Note that $D$ does not have $L, \nabla$ indexes, since in RPAE corresponding values are equal [3, 5].

**3.** We performed calculations using computing codes system ATOM-M [7]. Tables 1-4 collect the results for total $S$ and partial $S_i$ sums, defined by (1) and (2). We observe a remarkable feature: the partial sums $S_i$ are essentially different from, contrary to the wide spread believe, the number of electrons in the respective subshell $N_i$. The difference takes place not only in RPAE but also in HF, thus signaling redistributing of oscillator strength already on the one-electron HF level.

Note that after performing summation over all $i$, we obtain $S_{RPAE}^A < N_A$ - the total number of electron in atom A. The difference monotonic increase with atomic number growth, from 0.95 for Ar, to 2.38 for Pd, 4.56 for Xe, and reaches 9.23 for the heaviest considered atom Ra. These differences characterize the contributions to (1) of the cross-sections long "tails" that are beyond upper limits of our numeric integrations and of neglected discrete excitation levels, since we include only four of them for each subshell. However, this certainly does not affect the redistribution of the partial $S_i$ that is the main result of this Letter.

In most of the considered cases, the more electrons has a subshell, the bigger is the surplus $\Delta_i = S_i - N_i$ that goes from low-electron subshells due to intershell interaction. The very fact that this redistribution is a manifestation of the intershell interaction is easy to understand. Indeed, in absence of this interaction relation $S_{i,RPAE} = N_i$ becomes valid for any subshell. For equal numbers of $N_i$ of a given atom, $\Delta_i$ increases with growth of the principal quantum number. These tendencies manifest themselves already in Table 1 for Ar: 3p gains 1.8, while 2p – only 0.99. The "losers" are all *s*-subshells. In Pd the ranks of "losers" include also *p*-subshell while the "winners" are *d*-subshells, with $\Delta_{3d} = 4.19$ and $\Delta_{4d} = 5.32$. In Xe all subshells became losers, except 3d, 4d and outer 5p, with $\Delta_{3d} = 2.31$, $\Delta_{4d} = 4.82$, and $\Delta_{5p} = 2.73$. We considered also Ra. This is a very heavy element and relativistic calculations for it are necessary. However, we added non-relativistic results for completeness. Here "losers" are all but 4f, 5d and the outer 6p and even 7s subshells with $\Delta_{4f} = 10.58$, $\Delta_{5d} = 5.45$, $\Delta_{6p} = 2.92$ and $\Delta_{7s} = 0.39$, respectively.

In studies of multielectron atoms that require complex numeric calculations it is hard to explain results pure qualitatively, since the interplay of a number of different tendencies affect them. It is known, however that the cross-sections of subshells with small angular momentum $l = 0;1$, decreases with $\omega$ growth much slower than that with $l = 2;3$. The account of intershell interaction increases at high $\omega$ the cross sections of $l = 2;3$ subshells prominently [8]. In addition, one has to note that the interelectron interaction role is bigger when the relative role of nuclear charge is smaller, i.e. in outer subshells with many electrons and big principal quantum numbers. For them $\Delta_i$ are the biggest, as Tables 1-4 demonstrate.

**4.** The finding of this Letter surprises. Only here we demonstrate for the first time that intershell interaction affects prominently the absolute photoionization cross-section in a very broad frequency range, on the partial sum rule level. It demonstrates previously unnoted overall sufficiently strong intershell or interference interaction in atoms. Before, such interaction was considered as a rather specific feature that is effective in relatively narrow $\omega$ regions only in few outer shells, and near so-called Giant resonances of the intermediate subshells [9, 10]. No doubt that similar to presented here is the situation for all the atoms of the Periodic table and atom-like objects, e.g. clusters, fullerenes, endohedral – in each system with distinctive electronic shell structure.

It would be very interesting to perform experimental investigation aiming to demonstrate the prominent violation of the partial sum rules. This is not a simple task, having in mind that for each subshell *i* the measurements must be performed in a broad $\omega$ region in coincidence with creation of only *i* vacancy. However, such an experiment would be of great importance for the understanding of electronic structure of atoms and atom-like formations.

**Table 1.** Partial and total sums $S_{i,HF}^{L}$, $S_{i,HF}^{\nabla}$, $S_{i,RPAE}$, $\Delta_i \equiv S_{i,RPAE} - N_i$, and $\sum_{\leq i} S_{i,RPAE}$ of Ar atom.

| Ar, $N=Z=$**18** | Subshell $i$ | $N_i$ | $S_{i,HF}^{L}$ | $S_{i,HF}^{\nabla}$ | $S_{i,RPAE}$ | $\Delta_i$ | $\sum_{\leq i} S_{i,RPAE}$ |
|---|---|---|---|---|---|---|---|
| 1 | 1s | 2 | 0.80 | 0.78 | 0.80 | -1.20 | 0.80 |
| 2 | 2s | 2 | 1.25 | 1.19 | 1.00 | -1.00 | 1.80 |
| 3 | 2p | 6 | 7.420 | 6.29 | 7.00 | +0.99 | 8.80 |
| 4 | 3s | 2 | 0.612 | 0.53 | 0.46 | -1.54 | 9.26 |
| 5 | 3p | 6 | 10.41 | 5.84 | 7.80 | +1.8 | $S_{RPAE}$=**17.06** |

**Table 2.** Partial and total sums $S_{i,HF}^{L}$, $S_{i,HF}^{\nabla}$, $S_{i,RPAE}$, $\Delta_i \equiv S_{i,RPAE} - N_i$, and $\sum_{\leq i} S_{i,RPAE}$ of Pd atom.

| Pd, $N=Z=$**46** | Subshell $i$ | $N_i$ | $S_{i,HF}^{L}$ | $S_{i,HF}^{\nabla}$ | $S_{i,RPAE}$ | $\Delta_i$ | $\sum_{\leq i} S_{i,RPAE}$ |
|---|---|---|---|---|---|---|---|
| 1 | 1s | 2 | 0.59 | 0.59 | 0.59 | -1.41 | 0.59 |
| 2 | 2s | 2 | 1.00 | 1.00 | 1.00 | -1.00 | 1.59 |
| 3 | 2p | 6 | 3.36 | 3.2 | 3.43 | -2.57 | 5.02 |
| 4 | 3s | 2 | 1.02 | 0.96 | 0.88 | -1.12 | 5.90 |
| 5 | 3p | 6 | 5.05 | 4.15 | 4.13 | -1.87 | 10.03 |
| 6 | 3d | 10 | 15.75 | 12.95 | 14.19 | +4.19 | 24.22 |
| 7 | 4s | 2 | 0.755 | 0.66 | 0.67 | -1.33 | 24.89 |
| 8 | 4p | 6 | 2.81 | 2.43 | 3.41 | -2.59 | 28.3 |
| 9 | 4d | 10 | 21.57 | 11.79 | 15.32 | +5.32 | $S_{RPAE}$ =**43.62** |

**Table 3.** Partial and total sums $S_{i,HF}^{L}$, $S_{i,HF}^{\nabla}$, $S_{i,RPAE}$, $\Delta_i \equiv S_{i,RPAE} - N_i$, and $\sum_{\leq i} S_{i,RPAE}$ of Xe atom.

| Xe, $N=Z=$**54** | Subshell $i$ | $N_i$ | $S_{i,HF}^{L}$ | $S_{i,HF}^{\nabla}$ | $S_{i,RPAE}$ | $\Delta_i$ | $\sum_{\leq i} S_{i,RPAE}$ |
|---|---|---|---|---|---|---|---|
| 1 | 1s | 2 | 0.50 | 0.50 | 0.50 | -1.50 | 0.50 |
| 2 | 2s | 2 | 0.98 | 0.97 | 0.93 | -1.07 | 1.43 |
| 3 | 2p | 6 | 3.59 | 3.45 | 3.57 | -2.43 | 5.00 |
| 4 | 3s | 2 | 1.05 | 1.01 | 0.92 | -1.08 | 5.92 |
| 5 | 3p | 6 | 4.34 | 4.15 | 4.17 | -1.83 | 10.09 |
| 6 | 3d | 10 | 13.30 | 11.30 | 12.31 | +2.31 | 22.40 |

| 7 | 4s | 2 | 0.75 | 0.66 | 0.68 | -1.32 | 23.08 |
| 8 | 4p | 6 | 2.35 | 2.07 | 2.28 | -3.72 | 25.36 |
| 79 | 4d | 10 | 18.56 | 11.93 | 14.82 | +4.82 | 40.18 |
| 10 | 5s | 2 | 0.33 | 0.26 | 0.53 | -1.47 | 40.71 |
| 11 | 5p | 6 | 12.65 | 6.18 | 8.73 | +2.73 | $S_{\text{RPAE}}$ =**49.44** |

**Table 4.** Partial and total sums $S_{i,HF}^L$, $S_{i,HF}^\nabla$, $S_{i,RPAE}$, $\Delta_i \equiv S_{i,RPAE} - N_i$, and $\sum_{\leq i} S_{i,RPAE}$ of Ra atom.

| Ra, N=Z=**88** | Subshell $i$ | $N_i$ | $S_{i,HF}^L$ | $S_{i,HF}^\nabla$ | $S_{i,RPAE}$ | $\Delta_i$ | $\sum_{\leq i} S_{i,RPAE}$ |
|---|---|---|---|---|---|---|---|
| 1 | 1s | 2 | | | 0.05 | -1.50 | 0.05 |
| 2 | 2s | 2 | 0.243 | 0.24 | 0.23 | -1.77 | 0.28 |
| 3 | 2p | 6 | 1.04 | 1.02 | 1.05 | -0.95 | 1.34 |
| 4 | 3s | 2 | 0.57 | 0.56 | 0.52 | -1.48 | 1.86 |
| 5 | 3p | 6 | 2.5 | 2.44 | 2.48 | -3.52 | 4.34 |
| 6 | 3d | 10 | 7.25 | 6.77 | 7.45 | -2.55 | 11.79 |
| 7 | 4s | 2 | 0.84 | 0.78 | 0.76 | -1.24 | 12.54 |
| 8 | 4p | 6 | 3.14 | 2.97 | 3.04 | -2.96 | 15.58 |
| 9 | 4d | 10 | 8.05 | 7.64 | 8.41 | -1.59 | 23.99 |
| 10 | 4f | 14 | 27.57 | 21.88 | 24.58 | +10.58 | 48.57 |
| 11 | 5s | 2 | 0.73 | 0.66 | 0.67 | -1.33 | 49.25 |
| 12 | 5p | 6 | 2.08 | 1.89 | 2.16 | -3.84 | 51.41 |
| 13 | 5d | 10 | 19.44 | 12.12 | 15.45 | +5.45 | 66.86 |
| 14 | 6s | 2 | 0.367 | 0.303 | 0.60 | -1.40 | 67.46 |
| 15 | 6p | 6 | 6.92 | 3.54 | 8.92 | +2.92 | 76.38 |
| 16 | 7s | 2 | 0.25 | 0.25 | 2.39 | +0.39 | $S_{\text{RPAE}}$ =**78.77** |